\documentclass[fleqn,twoside,twocolumn,nofootinbib,showkeys]{revtex4} 
\usepackage[sec,doi]{ujp_UTF8} 

\begin{document}
\title[Structure of hypernucleus $_{\Lambda}^{7}$Li within microscopic three-cluster model]
{STRUCTURE OF HYPERNUCLEUS $_{\Lambda}^{7}$Li WITHIN MICROSCOPIC THREE-CLUSTER MODEL}%
\author{N.~Kalzhigitov}
\affiliation{Al-Farabi Kazakh National University, Almaty, Kazakhstan}
\address{71 al-Farabi Avenue, Almaty, 050040, Kazakhstan}
\email{knurto1@gmail.com}
\author{S.~Amangeldinova}
\affiliation{Al-Farabi Kazakh National University, Almaty, Kazakhstan}
\address{71 al-Farabi Avenue, Almaty, 050040, Kazakhstan}
\author{V. S.~Vasilevsky}
\affiliation{Bogolyubov Institute for Theoretical Physics, Kyiv, Ukraine}
\address{14b, Metrolohichna Str., Kyiv 03143, Ukraine}%
\udk{539.1}  \razd{\seci}
\razd{\seci}

\autorcol{N.\hspace*{0.7mm}Kalzhigitov, S.\hspace*{0.7mm}Amangeldinova, V.S.\hspace*{0.7mm} Vasilevsky}

\setcounter{page}{1}

\begin{abstract}
The structure of bound and resonance states of the hypernucleus $_{\Lambda}^{7}$Li is studied within a three-cluster model. This nucleus is considered a three-cluster structure consisting of $^{4}$He, a deuteron, and a lambda hyperon. The chosen three-cluster configuration allows us to describe more accurately the structure of hypernucleus $_{\Lambda}^{7}$Li and  the dynamics of different processes that involve interactions of lightest nuclei and hypernuclei. The main goal of the present investigations is to find resonance states in the three-cluster continuum of $_{\Lambda}^{7}$Li and determine their nature.
A set of narrow resonance states is detected at the energy range 	0$<E\leq$2 MeV above the three-cluster threshold $^{4}$He+$d$+$\Lambda$.
\end{abstract}


\maketitle

\section{Introduction}

The physics of hypernuclear systems has a relatively long and intriguing history. The main stages of this history are discussed in detail in Refs. \cite{2005NuPhA.754....3D, 2005NuPhA.754...14D}. There is a large number of experimental laboratories in the world that are trying to obtain new information on the structure of hypernuclei and their features. The geography of these laboratories and the main experimental methods they used to reveal the peculiarities of hypernuclei are presented in detail in a recently published review \cite{2025arXiv250600864C}. The main characteristics of the light hypernuclei are collected in the specialized database of hypernuclei  \cite{ChartHyperN2021}. This site displays the energies of bound states and dominant decay channels of hypernuclei.

Available experimental data stimulate a large number of theoretical investigations aimed at explaining the obtained experimental data and at predicting new features of hypernuclear systems and their interaction. Different theoretical models have been applied to study hypernuclear systems. Among them are the shell models, which described the p- and sd-shells  hypernuclei \cite{1978AnPhy.113...79G,
2012NuPhA.881..298M}, mean-field models \cite{1993PhRvC..48..889G,
2001PhRvC..64d4301V}, ab initio no-core shell models
\cite{2014PhRvL.113s2502W, 2020EPJA...56..301L}, as well as numerous
realizations of the cluster model \cite{1983PThPh..70..189M,
1985PThPS..81...42M, 2009PrPNP..63..339H,
2012FBS....53..189H, Nesterov:2021gcp,
2021NuPhA101622325N}. Usually, these investigations are devoted to studying the spectra of bound states of hypernuclei. Some of these investigations, carried out within cluster models, also considered resonance states only in the two-cluster continuum.

Our attention was attracted by the $_{\Lambda}^{7}$Li hypernucleus. An
interesting feature of the $_{\Lambda}^{7}$Li hypernucleus is that the spectrum consists of four bound states, exceeding the number of bound states in the ordinary ${}^{7}$Li nucleus, which has only two bound states. There is a lack of experimental and theoretical information about resonance states in this and other hypernuclei, which decay into two or three clusters. We wish to fill this gap by employing a three-cluster microscopic model. Thus, our main aim is to detect three-cluster resonance states in the  $_{\Lambda}^{7}$Li.

In the present work, we study the structure of both bound and resonance states of the $_{\Lambda}^{7}$Li hypernucleus. As we are mainly interested in the investigations of three-cluster resonance states, we adopted a three-cluster model, which was formulated in Ref. \cite{2001PhRvC..63c4606V} and was designed to study the decay of light nuclei into three fragments (clusters). This method has been successfully applied to study the structure of light atomic nuclei and especially resonance states in the three-cluster continuum of these nuclei  \cite{2010PPN....41..716N,
2012PhRvC..85c4318V, 2017PhRvC..96c4322V,
2018PhRvC..98b4325V}. Within the adopted model, the $_{\Lambda}^{7}$Li hypernucleus is considered as a three-cluster configuration, $^{4}$He+$d$+$\Lambda$. The interaction of six nucleons, split into two clusters, is modeled by a semi-realistic nucleon-nucleon potential, and the sum of nucleon-hyperon potentials determines the interaction of the lambda hyperon with two clusters. 

Our paper is organized as follows. In Sec. \ref{Sec:Method}, we explain the key elements of our model. The main results are presented in Sec.
\ref{Sec:Results}. Properties of bound states are considered in Sec. \ref{Sec:BoundSts}. A detailed discussion of the nature of resonance states is
given in Sec. \ref{Sec:ResonansSts}. We close the paper by summarizing the obtained results in Sec. \ref{Sec:Conclusions}.

\section{Method AMHHB \label{Sec:Method}}

We shortly present the essence of the three-cluster microscopic model which is usually referred as the AMHHB model, it means the model which uses the hyperspherical harmonics basis (HHB).

We start with the seven-particle system (six nucleons and one lambda hyperon) described by a microscopic Hamiltonian. Then we reduce it to an effective three-body problem by splitting  the seven particles into three groups (clusters). Then we assume that we know  wave functions describing the internal structure of each cluster with acceptable precision. Based on these assumptions, the wave function of $_{\Lambda}^{7}$Li is represented as%
\begin{eqnarray}
	\Psi_{J} &  =&\sum_{L,S}\widehat{\mathcal{A}}\left\{  \left[  \Phi_{1}\left(
	^{4}\text{He},S_{1}\right)  \Phi_{2}\left(  d,S_{2}\right)  \Phi_{3}\left(
	\Lambda,S_{3}\right)  \right]  _{S}\right.  \label{eq:M001}\\
	&  \times & \left.  \psi_{LSJ}\left(  \mathbf{x,y}\right)  \right\}
	_{J},\nonumber
\end{eqnarray}
where $\Phi_{1}\left(  ^{4}\text{He},S_{1}\right)  $ is the wave function of
an alpha particle, $\Phi_{2}\left(  d,S_{2}\right)  $ is the wave function of
a deuteron. As the lambda hyperon is considered a structureless particle, then
$\Phi_{3}\left(  \Lambda,S_{3}\right)  $ represents the spin part of the lambda hyperon function. The antisymmetrization operator $\widehat
{\mathcal{A}}$ permutes only nucleons and thus makes an antisymmetric wave function of $^{6}$Li, which is considered as a two-cluster system $^{4}$He+$d$.
Two Jacobi vectors $\mathbf{x}$ and $\mathbf{y}$ are used to determine the relative position of the clusters in the space. In what follows, the first vector $\mathbf{x}$ connects the center of masses of
$^{4}$He and the deuteron. In contrast, the second Jacobi vector $\mathbf{y}$ determines  the relative position of the lambda hyperon regarding the center of mass of $^{6}$Li.

To present the wave function of a three-cluster system (\ref{eq:M001}), we use the
LS coupling scheme. In this scheme, the total spin $S$ is a vector sum of the
individual spins of clusters. As the spin of $^{4}$He is equal to zero, the total spin of $_{\Lambda}^{7}$Li is the vector sum of the deuteron
spin ($S_{2}$=1) and the spin of the lambda hyperon ($S_{3}$=1/2). Thus, the total
spin of $_{\Lambda}^{7}$Li can be $S$=1/2 or $S$=3/2. Within the
present model, the total orbital momentum $L$ is the vector sum of the partial
orbital momenta of the relative motion of clusters (they will be introduced
later), and the total momentum $J$ is the vector sum of the total orbital
momentum $L$ and total spin $S$.

The wave function $\psi_{LSJ}\left(  \mathbf{x,y}\right)  $ of the relative motion of
clusters has to be determined by solving the Schr\"{o}dinger equation, 
	which is projected onto a three-cluster system and involves the selected
	nucleon-nucleon and nucleon-hyperon potentials. Note that the wave function
$\psi_{LSJ}\left(  \mathbf{x,y}\right)  $ depends on six variables presented
by the Jacobi vectors $\mathbf{x}$ and $\mathbf{y}$. Thus, we need
to introduce six quantum numbers to classify the states of a three-cluster system.
By using angular orbital momentum reduction, we represent this function as
\begin{equation}
	\psi_{LM_{L}}\left(  \mathbf{x},\mathbf{y}\right)  \Rightarrow\sum_{\lambda
		,l}\psi_{\lambda,l;L}\left(  x,y\right)  \left\{  Y_{\lambda}\left(
	\widehat{\mathbf{x}}\right)  Y_{l}\left(  \widehat{\mathbf{y}}\right)
	\right\}  _{LM_{L}},\label{eq:M002}%
\end{equation}
where $\widehat{\mathbf{x}}$ and $\widehat{\mathbf{y}}$ are unit vectors, and
$\lambda$ and $l$ are the partial angular momenta associated with vectors
$\mathbf{x}$ and $\mathbf{y}$, respectively. With such a reduction, we define four
quantum numbers $\lambda$, $l$, $L$ and $M$. Within the present model, the
total orbital momentum $\overrightarrow{L}$ is a vector coupling of partial
orbital momenta $\overrightarrow{L}=\overrightarrow{\lambda}+\overrightarrow
{l}$.

Wave functions of inter-cluster motion $\psi_{\lambda,l;L}\left(  x,y\right)
$ obey an infinite set of two-dimensional integro-differential equations. To
solve this set of equations, we employ the hyperspherical coordinates and
hyperspherical harmonics. There are several equivalent sets of the
hyperspherical harmonics in the literature, which involve different sets of
hyperspherical coordinates. We select the hyperspherical harmonics in the
form suggested by Zernike and Brinkman in Ref. \cite{kn:ZB}. This form of the
hyperspherical harmonics is fairly simple and does not require bulky
analytical and numerical calculations. To construct the Zernike - Brinkman
hyperspherical harmonics, we need to introduce two hyperspherical coordinates
$\rho$ and $\theta$ instead of scalar coordinates $x$ and $y$. The first
coordinate $\rho$ is a hyperspherical radius
\begin{equation}
	\rho=\sqrt{x^{2}+y^{2}} , \label{eq:M005A}%
\end{equation}
and the second coordinate $\theta$ is a hyperspherical angle%
\begin{equation}
	\theta=\arctan\left(  \frac{x}{y}\right)  . \label{eq:M005B}%
\end{equation}
With a fixed value of $\rho$, this angle determines the relative length of the
vectors $\mathbf{x}$ and $\mathbf{y}$%
\begin{equation}
	x=\cos\theta,\quad y=\rho\sin\theta. \label{eq:M006}%
\end{equation}
One can see that the hyperradius $\rho$\ determines the size of the triangle
$\theta$ which connects the centers of mass of three clusters, and the
hyperangle $\theta$ determines its shape.

In new coordinates, wave function (\ref{eq:M001}) can be presented as%
\begin{eqnarray}
	&&\Psi_{J}  =  \nonumber\\  &=&\sum_{L,S}\sum_{K,\lambda,l}\widehat{\mathcal{A}}\left\{
	\left[  \Phi_{1}\left(  ^{4}\text{He},S_{1}\right)  \Phi_{2}\left(
	d,S_{2}\right)  \Phi_{3}\left(  \Lambda,S_{3}\right)  \right]  _{S}\right.
	\nonumber\\
	&  \times & \left.  R_{c}\left(  \rho\right)  \mathcal{Y}_{c}\left(
	\Omega\right)  \right\}  _{J},
\end{eqnarray}
where $K$ is the hypermomentum, and $\mathcal{Y}_{c}\left(  \Omega\right)  $
stands\ for the product
\begin{equation}
	\mathcal{Y}_{c}\left(  \Omega\right)  =\chi_{K}^{\left(  \lambda,l\right)
	}\left(  \theta\right)  \left\{  Y_{\lambda}\left(  \widehat{\mathbf{x}%
	}\right)  Y_{l}\left(  \widehat{\mathbf{y}}\right)  \right\}  _{LM_{L}}
	\label{eq:M012}%
\end{equation}
and represents a hyperspherical harmonic for a three-cluster channel
\begin{equation}
	c=\left\{  K,\lambda,l,L\right\}  . \label{eq:M013}%
\end{equation}
The hyperspherical harmonic $\mathcal{Y}_{c}\left(  \Omega\right)  $ is a
function of five angular variables $\Omega=\left\{  \theta,\widehat
{\mathbf{x}},\widehat{\mathbf{y}}\right\}  $. The definition of all components of
the hyperspherical harmonic $\mathcal{Y}_{c}\left(  \Omega\right)  $ can be
found, for example, in Ref. \cite{2001PhRvC..63c4606V}. Being a complete
basis, the hyperspherical harmonics account for any shape of the three-cluster
triangle and its orientation. Thus, they account for all possible modes of
relative motion of  the three interacting clusters.

And the final step toward the numerical investigation of the three-cluster system is in our hands. To simplify solving a set of integro-differential equations for
hyperradial wave functions $R_{c}\left(  \rho\right)  $, we expand them over
the full set of oscillator functions $\Phi_{n_{\rho},K}\left(  \rho,b\right)
$
\begin{equation}
	R_{c}\left(  \rho\right)  =\sum_{n_{\rho},c}C_{n_{\rho},c}\Phi_{n_{\rho}%
		,K}\left(  \rho,b\right)  . \label{eq:M021}%
\end{equation}
As a result, a set of integro-differential equations is reduced to a
system of linear algebraic equations%
\begin{equation}
	\sum_{\widetilde{n}_{\rho},\widetilde{c}}\left[  \left\langle n_{\rho
	},c\left\vert \widehat{H}\right\vert \widetilde{n}_{\rho},\widetilde
	{c}\right\rangle -E\left\langle n_{\rho},c|\widetilde{n}_{\rho},\widetilde
	{c}\right\rangle \right]  C_{\widetilde{n}_{\rho},\widetilde{c}}=0,
	\label{eq:M022}%
\end{equation}
where $\left\langle n_{\rho},c\left\vert \widehat{H}\right\vert \widetilde
{n}_{\rho},\widetilde{c}\right\rangle $ is the matrix elements of the
three-cluster Hamiltonian, and $\left\langle n_{\rho},c|\widetilde{n}_{\rho
},\widetilde{c}\right\rangle $ is the matrix elements of the norm kernel. The
oscillator function $\Phi_{n_{\rho},K}\left(  \rho,b\right)  $  (or
	more precisely, the radial part of the wave function of the six-dimensional oscillator)
is
\begin{eqnarray}
	\Phi_{n_{\rho},K}\left(  \rho,b\right)   &  =&\left(  -1\right)  ^{n_{\rho}%
	}\mathcal{N}_{n_{\rho},K}\label{eq:M023}\\
	&  \times& r^{K}\exp\left\{  -\frac{1}{2}r^{2}\right\}  L_{n_{\rho}}%
	^{K+3}\left(  r^{2}\right)  ,\nonumber\\
	r    = \rho/b, &\quad & \mathcal{N}_{n_{\rho},K}=b^{-3}\sqrt{\frac{2\Gamma\left(
			n_{\rho}+1\right)  }{\Gamma\left(  n_{\rho}+K+3\right)  }},\nonumber
\end{eqnarray}
and $b$ is an oscillator length.

System of equations (\ref{eq:M022}) can be solved numerically by imposing
restrictions on the number of hyperradial excitations $n_{\rho}$ and the
number $N_{ch}$ of hyperspherical channels $c_{1}$, $c_{2}$, \ldots,
$c_{N_{ch}}$. The diagonalization procedure is used to determine energies and
wave functions of the bound states. However, the proper boundary conditions
have to be implemented to calculate elements of the scattering $S$-matrix and
corresponding functions of the continuous spectrum. \ Boundary conditions for wave
functions of two- and three-body decays of a compound three-cluster system are
thoroughly discussed in Refs \cite{2001PhRvC..63c4606V,
2018PhRvC..97f4605V}.

To analyze wave functions of many-channel system, obtained by solving the set
of equations (\ref{eq:M022}), it is expedient to combine those oscillator wave
functions (\ref{eq:M023}), which belong to the oscillator shell with the total
number of oscillator quanta $N_{os}=2n_{\rho}+K$. It is then convenient to
numerate the oscillator shells by quantum number $N_{sh}$ ( = 0, 1, 2,
\ldots), which we determine as%
\[
N_{os}=2n_{\rho}+K=2N_{sh}+K_{\min},
\]
where $K_{\min}=L$ for normal parity states $\pi=\left(  -1\right)  ^{L}$ and
$K_{\min}=L+1$ for abnormal parity states $\pi=\left(  -1\right)  ^{L+1}$. In
what follows, we will study the weights $W_{sh}\left(  N_{sh}\right)  $  of oscillator wave functions of a fixed oscillator shell
$N_{sh}$ in the  wave function of bound or continuous spectrum
states. The weights $W_{sh}\left(  N_{sh}\right)  $ are determined as%
\begin{equation}
	W_{sh}\left(  N_{sh}\right)  =\sum_{n_{\rho},K\in N_{sh}}\left\vert
	C_{n_{\rho},c}\right\vert ^{2} \label{eq:M025}%
\end{equation}
and indicate whether the system under consideration is compact (the oscillator
shells with small values of $N_{sh}$ dominate) or relatively dispersed (the
oscillator shells with large values of $N_{sh}$ dominate).

\section{Results and discussions \label{Sec:Results}}

For a detailed investigation of bound and resonance states of $_{\Lambda
}^{7}$Li, we selected the Hasegawa-Nagata potential (HNP) \cite{potMHN1,
	potMHN2} and a nucleon-hyperon potential \cite{1994PThPS.117..361Y}, which is
usually called the YNG-NF potential. The oscillator length $b$, which is
the only free parameter of our model and which determines the distribution of
nucleons inside clusters $^{4}$He and $d$, is chosen to minimize the energy of the
three-cluster threshold $^{4}$He+$d$+$\Lambda$ and for the HNP, equals
$b$=1.357 fm. With this value of $b$, the bound state energy of $^{6}$Li
accounted for from the two-cluster threshold $^{4}$He+$d$ is $E$(1$^{+}$) =-1.431
MeV, which is close to the experimental value $E$(1$^{+}$) =-1.474 MeV.

Note that with the chosen three-cluster configuration $^{4}$He+$d$+$\Lambda$,
we have three two-cluster subsystems $^{4}$He+$d$, $d$+$\Lambda$ and $^{4}%
$He+$\Lambda$. They are of importance for the present calculations. We rely
on the results of Ref. \cite{Kalzhigitov2020Bul}, where the interaction of a deuteron
with an alpha particle was considered, and on the results of Ref.
\cite{Kalzhigitov2025}, where the interaction of a lambda hyperon with a deuteron and an
alpha particle was investigated in detail.

After the oscillator length $b$ and the nucleon-nucleon ($NN$) and
nucleon-hyperon ($N\Lambda$) potentials were selected, we need to fix another
two input parameters: the number of channels or number of hyperspherical
harmonics, and the number of hyper-radial excitations. We have to restrict
ourselves to a finite set of hyperspherical harmonics, which is determined
by the maximal value of the hyperspherical momentum $K_{\max}$. To describe
the positive parity states, we use all hyperspherical harmonics with the
hypermomentum $K$ $\leq K_{\max}$ = 12, and the negative parity states are
represented by the hyperspherical harmonics with $K$ $\leq K_{\max}$ = 11. These
numbers of hyperspherical harmonics allow us \ to describe a large number of
scenarios of the three-cluster decay. We also have  to restrict ourselves to
the number of hyper-radial excitations $n_{\rho}\leq$100. This number of
hyper-radial excitations allows us to reach the asymptotic region, where all
clusters are well separated, and the inter-cluster interaction induced by the
$NN$ or/and $N\Lambda$ potentials becomes negligibly small.

\subsection{Bound states \label{Sec:BoundSts}}

The spectrum of $_{\Lambda}^{7}$Li \ bound states, which is obtained with the HNP
and YNG potentials, is shown in Table \ref{Tab:Spectr7HLiHNP}. The energy of bound
states is reckoned from the three-cluster threshold $^{4}$He+$d$+$\Lambda$.
The energies of the deeply bound 1/2$^{+}$, 3/2$^{+}$ and 5/2$^{+} $, obtained
with our model, are very close to the experimental values. However, our model
generates the weakly bound 1/2$^{+}$ state, which is approximately 2.1 MeV underbound. The mass root-mean-square radii $R_{m}$ indicate that
the deeply bound states are compact states, with 2.0$<R_{m}<$2.2 fm, while the
weakly bound state is a very dispersed state with a large value of $R_{m}=$4.5 fm.%

	\begin{table}[tbp] \centering
		\noindent\caption{Spectrum of $^{7}_{\Lambda}$Li calculated with the HNP and YNG
			potentials}%
			\vskip3mm\tabcolsep4.5pt
		\noindent{\footnotesize\begin{tabular}
			[c]{|c|c|c|c|}\hline
			& \multicolumn{2}{|c|}{AM HHB} & Exp\\\hline
			$J^{\pi}$ & $E$, MeV & $R_{m}$, fm & $E$, MeV\\\hline
			1/2$^{+}$ & -7.060 & 2.183 & -7.094\\
			3/2$^{+}$ & -6.587 & 2.208 & -6.402\\
			5/2$^{+}$ & -4.856 & 2.036 & -5.043\\
			1/2$^{+}$ & -1.113 & 4.524 & -3.217\\\hline
			$^{4}$He$+d+\Lambda$ & 0.0 & - & 0.0\\\hline
		\end{tabular}}
		\label{Tab:Spectr7HLiHNP}%
\end{table}%

\begin{figure}
	[ptb]
	\begin{center}
		\includegraphics[
		width=0.48\textwidth
		]%
		{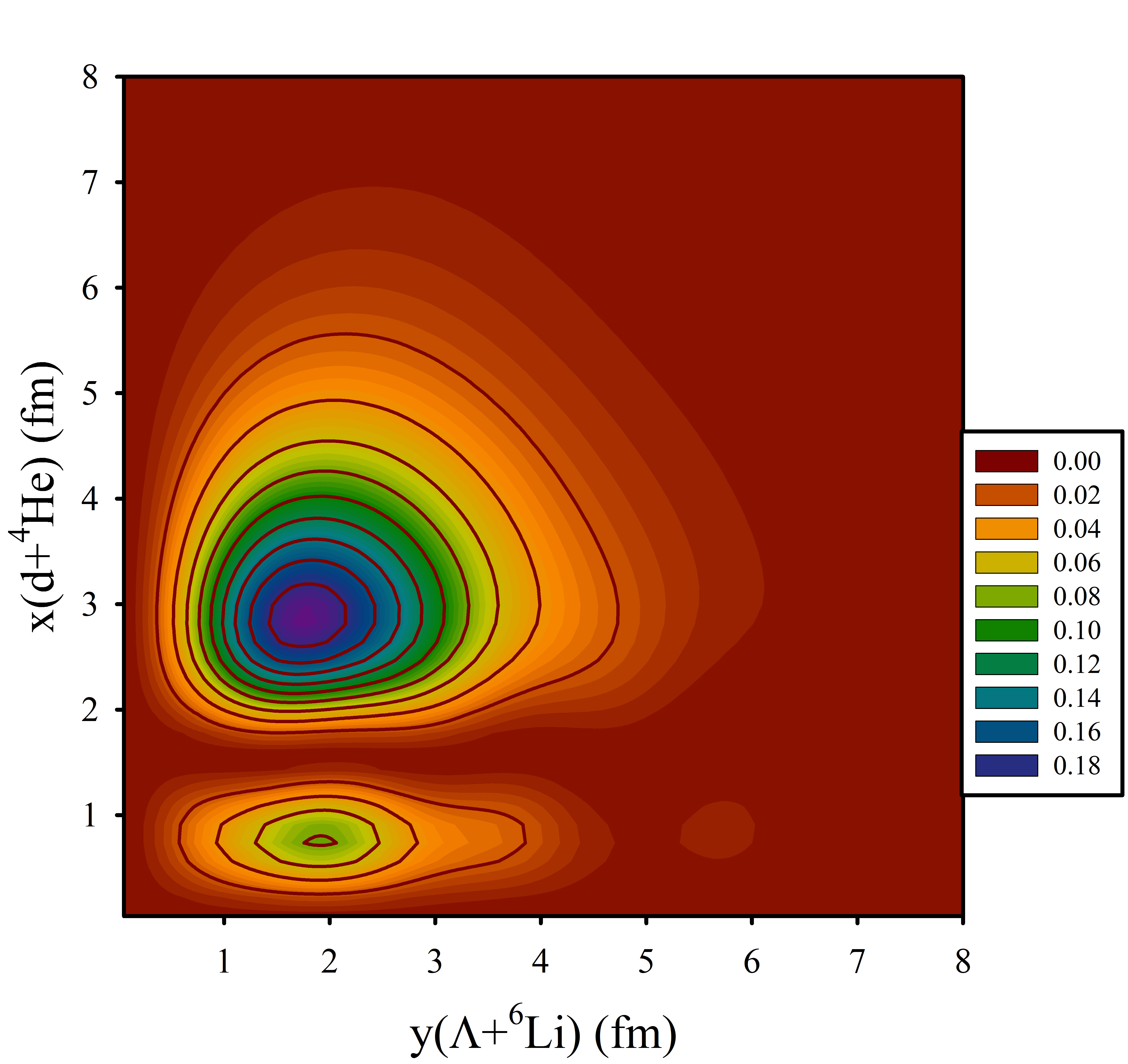}%
		\caption{Correlation function of the $_{\Lambda}^{7}$Li \ ground state as a
			function of distances $x$ and $y$. The length of vector $\mathbf{x}$
			determines distance between deuteron and alpha particle, and the length of
			vector $\mathbf{y}$ determines distance between lambda hyperon and $^{6}$Li}%
		\label{Fig:CorrFun7HLiGS}%
	\end{center}
\end{figure}

To understand the structure of bound states, we consider the correlation
functions $D\left(  x,y\right)  $, which are determined as
\[
D\left(  x,y\right)  =\left(  xy\right)  ^{2}\sum_{\lambda,l,L}\left\vert
\psi_{\lambda,l;L}\left(  x,y\right)  \right\vert ^{2}.
\]
Note that the correlation function determines the most probable geometry
(relative position) of three interacting clusters. In Fig.
\ref{Fig:CorrFun7HLiGS}, we display a correlation function for the ground state
of $_{\Lambda}^{7}$Li. The main peak of the correlation function corresponds
to the three-cluster configuration where the distance between the deuteron and
alpha particle is approximately 2.9 fm and the lambda particle is located close to
the center of mass of $^{6}$Li at a distance of 1.7 fm.

It is worthwhile noticing that the 3/2$^{-}$ ground state of the ordinary nucleus
$^{7}$Li, determined with the three-cluster configuration $^{4}$He+$d$+$n$ and
with the same input parameters, has a lower bound state energy (-11.24 MeV concerning the three-cluster threshold, while the bound state energy of $_{\Lambda
}^{7}$Li is -7.06 MeV) and it should be more compact than $_{\Lambda}^{7}%
$Li. However, the most probable distance between a deuteron and an alpha particle
is $x$ = 3.45 fm, and the distance between a neutron and $^{6}$Li is $y$=2.05 fm.
Such distances reflect that the ground state of $^{7}$Li is mainly
 the two-cluster configuration $^{3}$H+$^{4}$He, where
the valence neutron is very close to the deuteron. (See Ref.
 \cite{2009PAN....72.1450N} for details of such calculations.). Besides, the
Pauli principle plays an important role in the formation of the bound states of
$^{7}$Li. The antisymmetrization over all nucleons creates the Pauli forbidden
states, which are not observed in $_{\Lambda}^{7}$Li.%

\begin{figure}
	[ptb]
	\begin{center}
		\includegraphics[
		width=0.48\textwidth
		]%
		{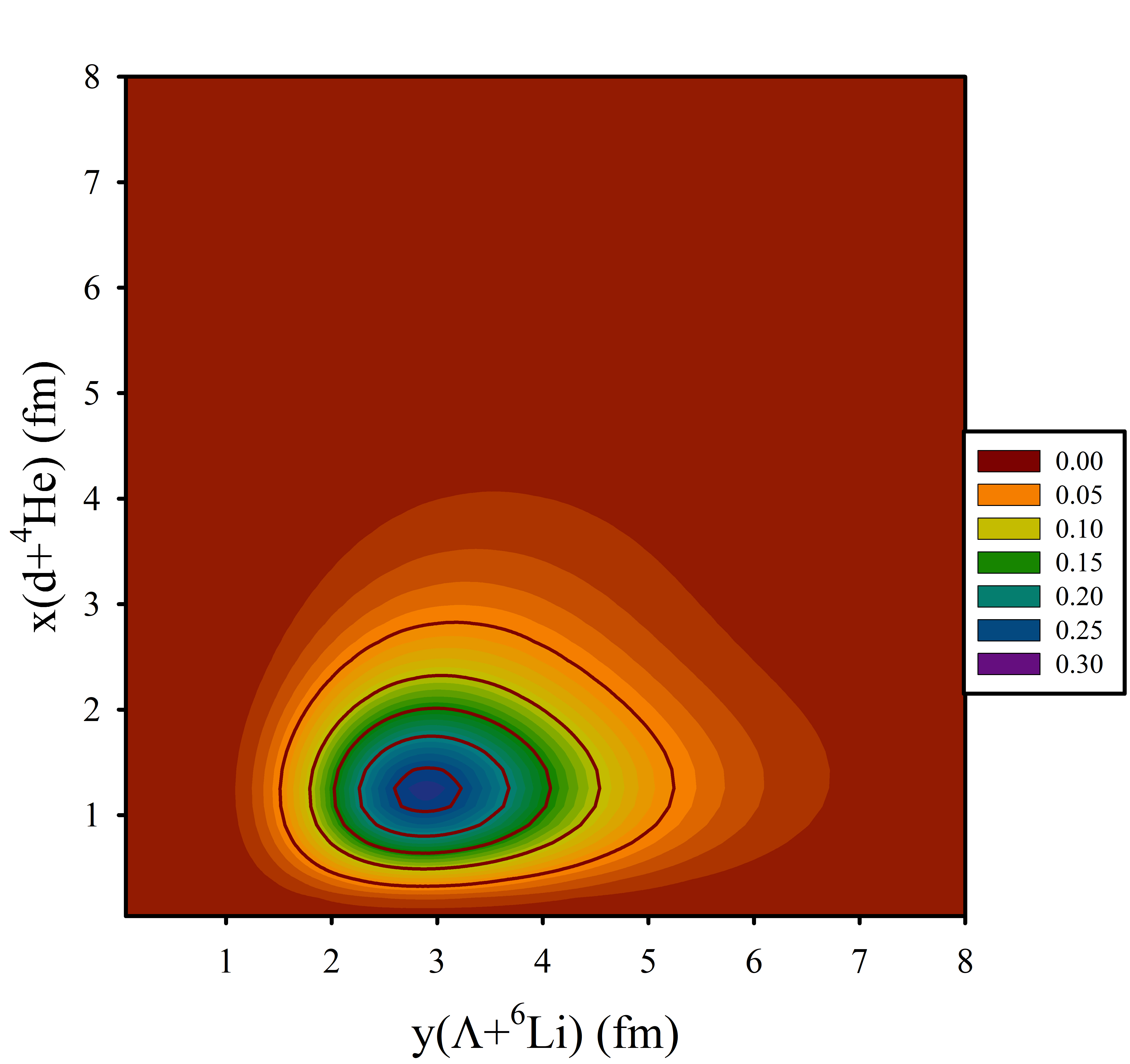}%
		\caption{Correlation function of the 5/2$^{+}$ excited state in $_{\Lambda
			}^{7}$Li}%
		\label{Fig:CorrFun7HLi52P}%
	\end{center}
\end{figure}

The correlation function for the 5/2$^{+}$ excited state of  $_{\Lambda}^{7}$Li
is shown in Fig. \ref{Fig:CorrFun7HLi52P}. Comparing correlation functions for
5/2$^{+}$ and 1/2$^{+}$ states, we see that in the 5/2$^{+}$ excited state, the
lambda hyperon is rather far from $^{6}$Li compared to the 1/2$^{+}$ states.
Besides, the distance between the deuteron and the alpha particle, forming $^{6}$Li,
is much smaller in the 5/2$^{+}$ state than in the 1/2$^{+} $ state. The peak
of the correlation function for the 5/2$^{+}$ state is located at $x$=1.27 fm
and $y$=2.95 fm. It means that in this state, the distance between the deuteron
and the alpha particle is almost two times smaller than in the 1/2$^{+}$ ground
state. For comparison, the distance between the lambda hyperon and $^{6}$Li is
approximately two times smaller.%

\begin{figure}
	[ptb]
	\begin{center}
		\includegraphics[
		width=0.48\textwidth
		]%
		{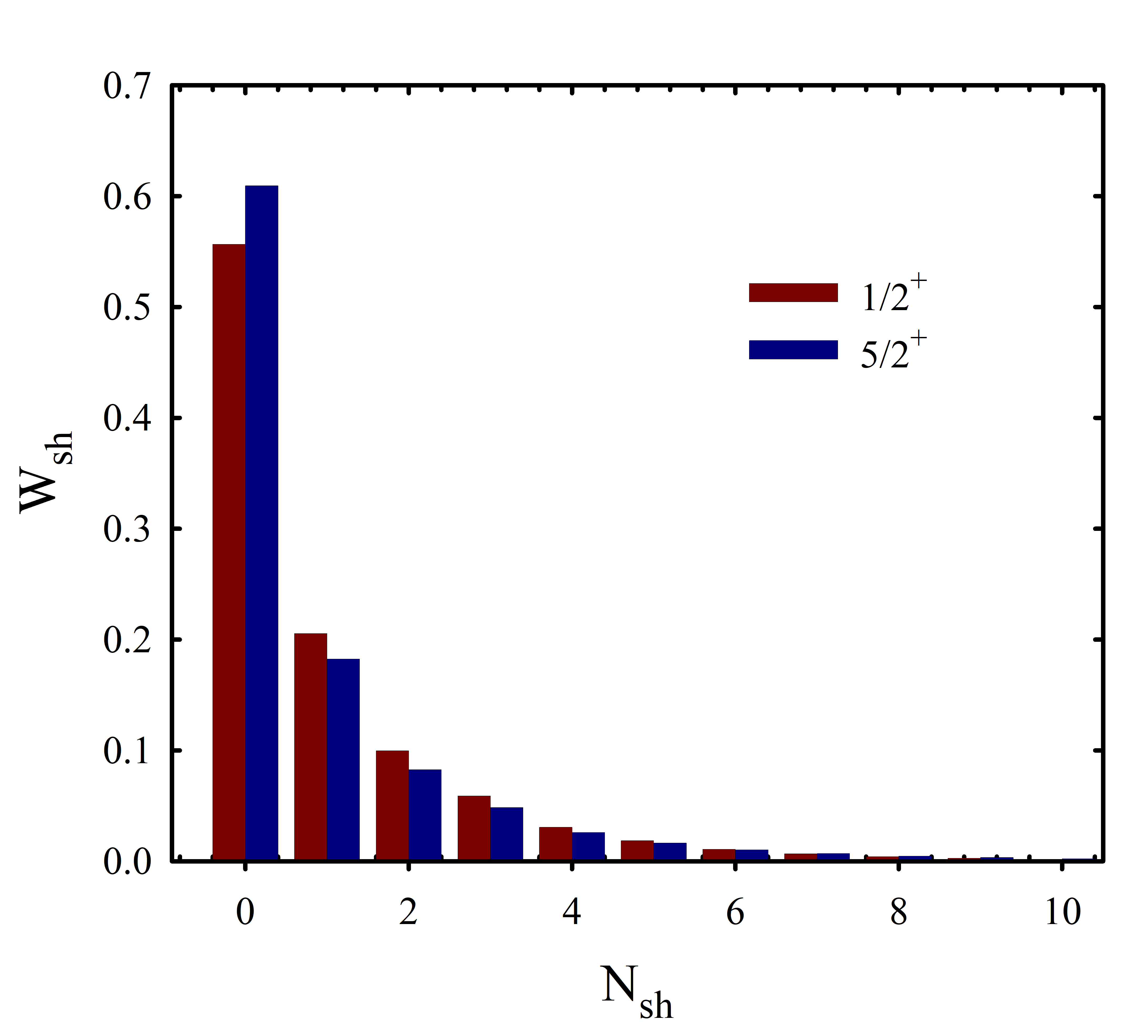}%
		\caption{Weights of different oscillator shells to the wave functions of
			1/2$^{+}$ and 5/2$^{+}$ states in $_{\Lambda}^{7}$Li. }%
		\label{Fig:ShellWeighs7HLit}%
	\end{center}
\end{figure}

Additional information about peculiarities of bound states of $_{\Lambda}^{7}%
$Li can be obtained by analyzing the weights of different oscillator shells in
wave functions of bound states. The definition of such quantities can be found in
Refs. \cite{2012PhRvC..85c4318V,  2017PhRvC..96c4322V,
 2018PhRvC..98b4325V}. In Fig. \ref{Fig:ShellWeighs7HLit}, the weights of
different oscillator shells $W_{sh}$ are displayed for the 1/2$^{+}$ ground
and first excited 5/2$^{+}$ states. The largest contribution of the lowest
oscillator shell $N_{sh}$=0 to the wave functions of interest indicates that the
lambda hyperon with a large probability ($>$50\%) can be found inside the nucleus $^{6}$Li.

\bigskip%

\begin{figure}
	[ptb]
	\begin{center}
		\includegraphics[
		width=0.48\textwidth
		]%
		{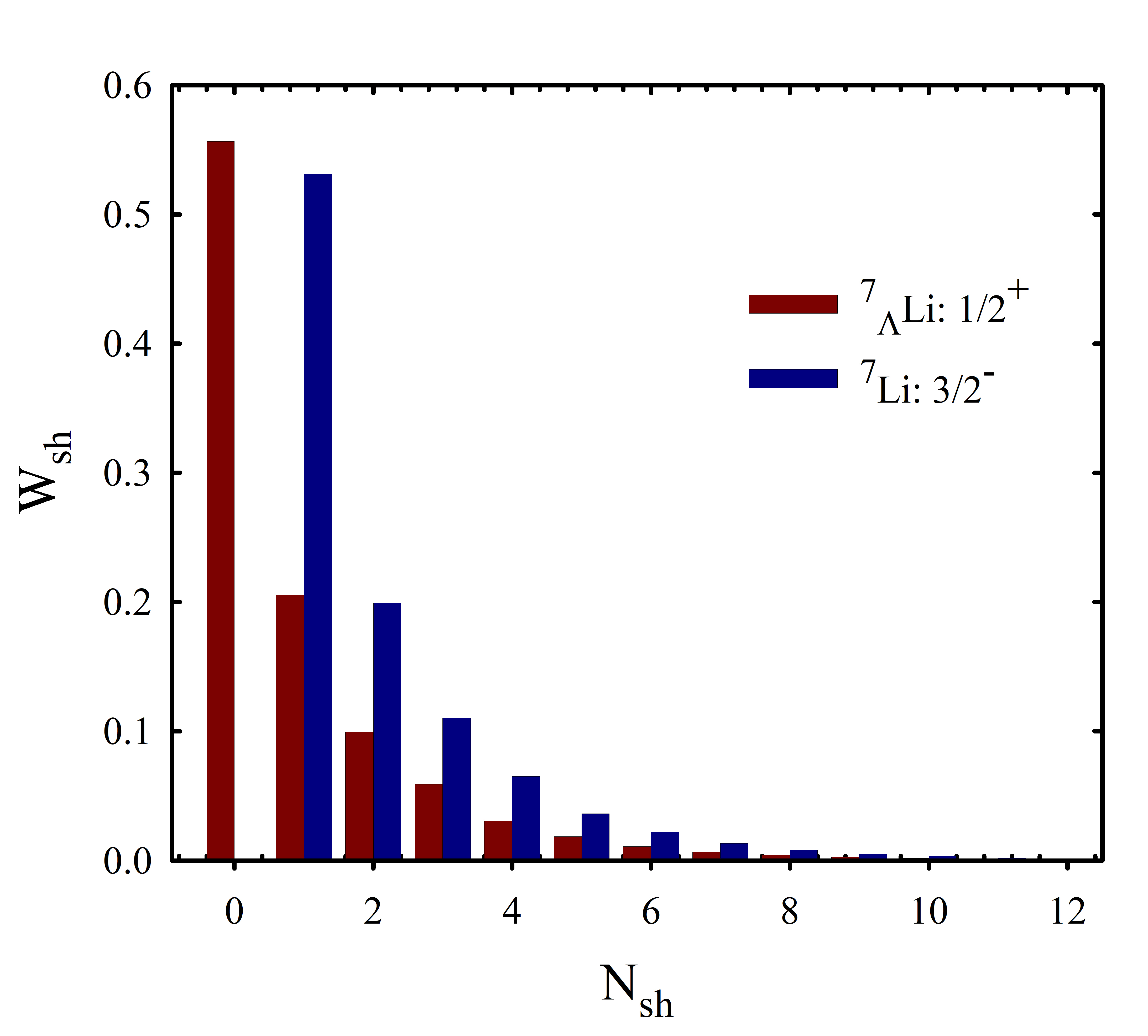}%
		\caption{Decomposition of wave functions of the $^{7}$Li and $_{\Lambda}^{7}%
			$Li ground states over oscillator shells. PP indicates effects of the Pauli
			principle.}%
		\label{Fig:ShellWeighs7Li7HLiGS}%
	\end{center}
\end{figure}

In Fig. \ref{Fig:ShellWeighs7Li7HLiGS}, we compare the structure of wave functions
of the ground states of $^{7}$Li and $_{\Lambda}^{7}$Li. Recall that the
 3/2$^{^{-}}$ state is the ground state of $^{7}$Li and the 1/2$^{+}$ state represents the ground state. One can see that the lowest shell
$N_{sh}$=0 gives zero contribution to the wave function of the $^{7}$Li ground
state. This shell describes the condensate of three clusters $^{4}$He, $d$, $n$,
and thus it is a forbidden shell for $^{7}$Li due to the Pauli principle.%

\subsection{Resonance states \label{Sec:ResonansSts}}

In Table \ref{Tab:3CResonans}, we collect information on resonance states of
$_{\Lambda}^{7}$Li determined in the three-cluster continuum $^{4}%
$He$+d+\Lambda$. The energies of resonance states are found in the energy range
from 0.2 to 2 MeV. It seems that the state 3/2$^{-}$ generates the largest
kinematical and Coulomb barrier, which resides in two resonance states with
relatively small total widths $\Gamma$. The ratio $\Gamma/E$  is used to
distinguish very narrow, narrow and relatively wide resonance states (see, for
example, Ref. \cite{2017PhRvC..96c4322V,  2018PhRvC..98b4325V}). It
indicates that the 5/2$^{-}$ resonance state with the energy $E$=0.290 MeV is
the narrowest resonance state (with the total width of 1.0 keV) in the
three-cluster continuum of $_{\Lambda}^{7}$Li and the 3/2$^{+}$ resonance
state with the energy $E$=1.604 MeV is the widest resonance state.%

	\begin{table}[tbp] \centering
		\noindent\caption{Parameters of resonance states found in three-cluster continuum of
			$^7_{\Lambda}$Li}%
		\vskip3mm\tabcolsep4.5pt
		\noindent{\footnotesize\begin{tabular}
			[c]{|l|l|l|l|}\hline
			$J^{\pi}$ & $E$, MeV & $\Gamma$, MeV & $\Gamma/E$\\\hline
			5/2$^{+}$ & 0.290 & 0.00104 & 3.6$\times$10$^{-7}$\\\hline
			1/2$^{-}$ & 1.545 & 0.264 & 0.171\\\hline
			3/2$^{-}$ & 1.043 & 0.279 & 0.267\\\hline
			3/2$^{-}$ & 1.551 & 0.249 & 0.161\\\hline
			1/2$^{+}$ & 1.520 & 0.547 & 0.360\\\hline
			3/2$^{+}$ & 1.604 & 0.696 & 0.434\\\hline
		\end{tabular}}
		\label{Tab:3CResonans}%
\end{table}%

\begin{figure}
	[ptb]
	\begin{center}
		\includegraphics[
		width=0.48\textwidth
		]%
		{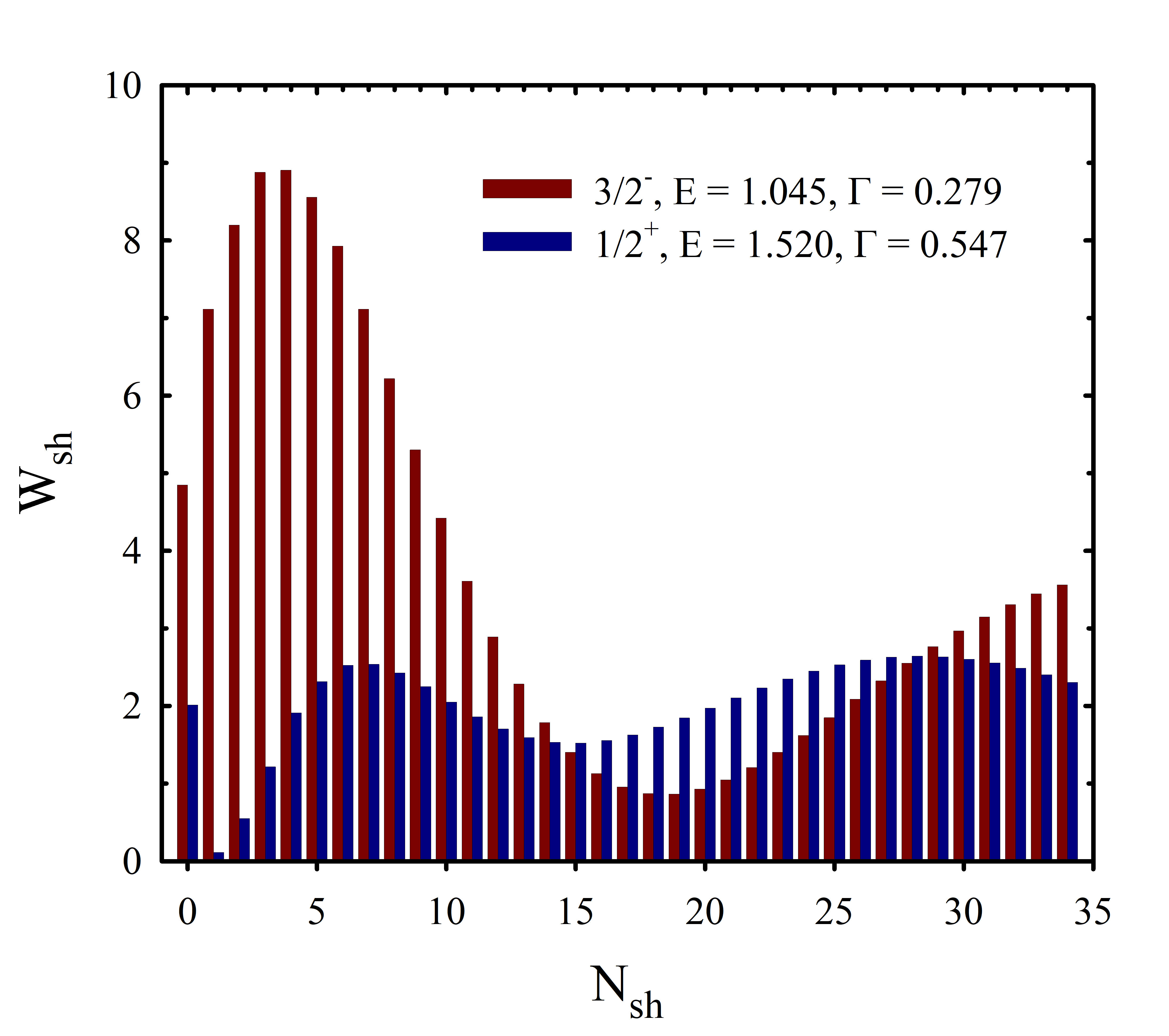}%
		\caption{Weights of wave function of different oscillator shell in wave
			functions of the 3/2$^{-}$ and 3/2$^{+}$ resonance states}%
		\label{Fig:ShellWeighs7HLiRS}%
	\end{center}
\end{figure}

To understand the nature of resonance states, it is expedient to analyze resonance wave functions. In Fig.
\ref{Fig:ShellWeighs7HLiRS}, the weights of different oscillator shells in wave
functions of the narrowest 3/2$^{-}$ states are compared with the weights of the widest 3/2$^{+}$ resonance state. The wave function of a narrow resonance
state has a large contribution of the oscillator shells with small values of
$N_{sh}$, namely, 0$\leq N_{sh}\leq$10.  It is necessary to recall that
oscillator wave functions of these shells describe the most compact
three-cluster configurations. The wave function of a rather wide resonance
state is spread over a large number of oscillator shells.%

\begin{figure}
	[ptb]
	\begin{center}
		\includegraphics[
		width=0.48\textwidth
		]%
		{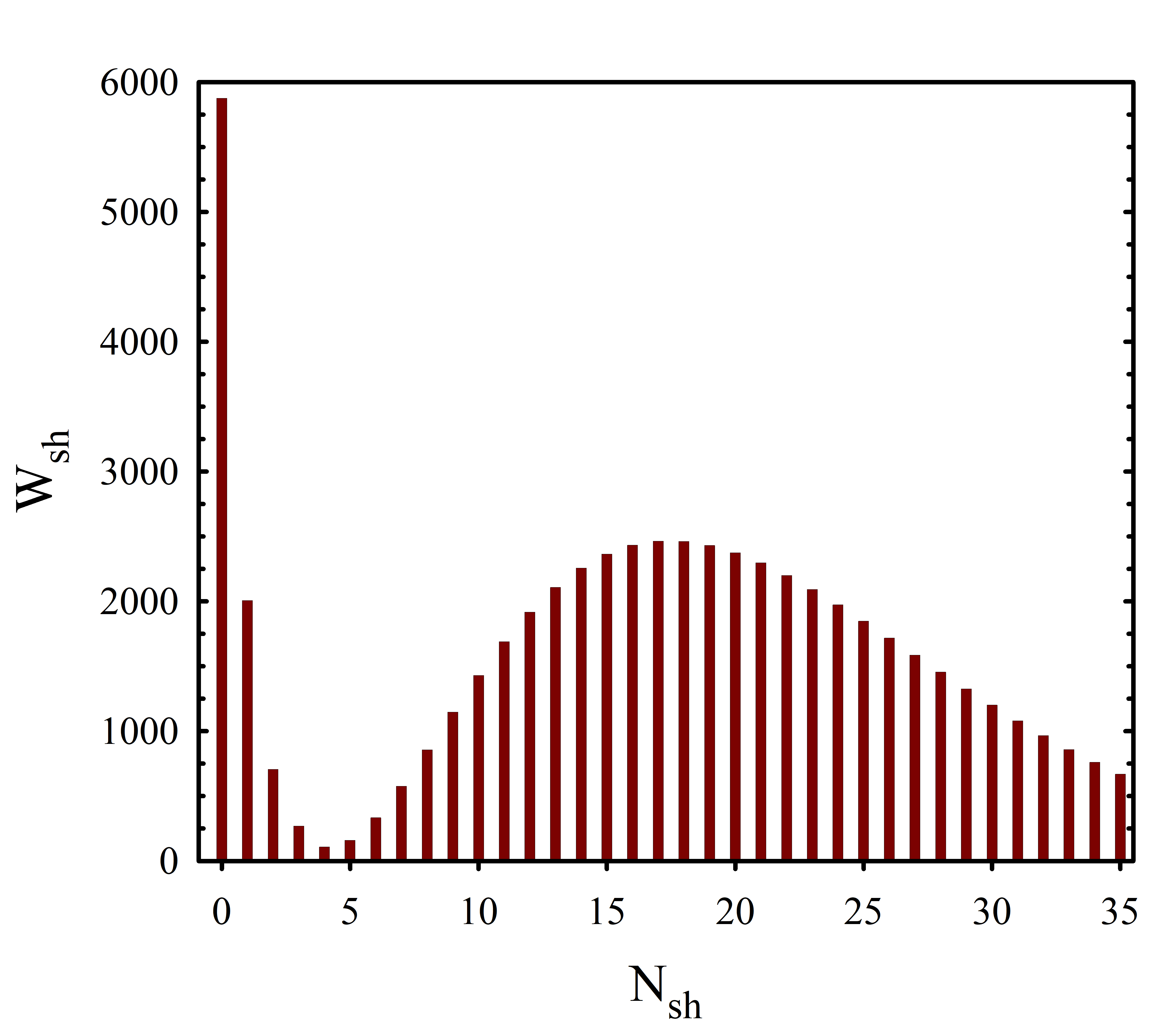}%
		\caption{Weights of different oscillator shells in the wave function of the
			super narrow 5/2$^{+}$ resonance state}%
		\label{Fig:ShellWeighs7HLi52PRS}%
	\end{center}
\end{figure}

The structure of the wave function of the very (super) narrow 5/2$^{+}$ resonance
is shown in Fig. \ref{Fig:ShellWeighs7HLi52PRS}. The weights of oscillator
shells in the wave function of the resonance state have much larger amplitudes.
Such enormous amplitudes have been observed for the long-lived Hoyle state in
$^{12}$C \cite{2012PhRvC..85c4318V} and for Hoyle-analogue states in some
light nuclei \cite{2018PhRvC..98b4325V}. It is necessary to note that 126
channels participate in the formation of the 5/2$^{+}$ continuous spectrum
states. And only one channel dominates in the decay (or formation) of the super-narrow 5/2$^{+}$ resonance state. This channel has quantum numbers:
$c=\left\{  K=2,l_{1}=0,l_{2}=2,L=2,S=1/2\right\}  $.

\bigskip\ %

\section{Conclusions \label{Sec:Conclusions}}

We have studied the structure of the $_{\Lambda}^{7}$Li hypernucleus  by
employing a three-cluster microscopic model, which allows us to study not only
bound states but three-cluster resonance states. We calculated energies and
wave functions of bound states of $_{\Lambda}^{7}$Li, and revealed those
channels that give the maximal contribution to the wave function of these
states. We also calculated the mass root-mean-square radii of the bound
states, which indicate that the hypernucleus $_{\Lambda}^{7}$Li is more
compact than the ordinary $^{7}$Li nucleus. It was shown that the present
model fairly good describes the bound states of $_{\Lambda}^{7}$Li. It was
also demonstrated that all but one bound states of $_{\Lambda}^{7}$Li are very
compact states with small values of the mass root-mean-square radius.
Correlation functions revealed the most probable relative position
(distribution) of clusters in coordinate space. Besides, the weights of the
functions of a fixed oscillator shell in the wave functions of the bound
states of $_{\Lambda}^{7}$Li unambiguously demonstrate that the lambda hyperon
can be located inside the nucleus ${}^{6}$Li with significant probability.

The microscopic model we employed involves hyperspherical harmonics to
numerate channels of a three-cluster system and to implement proper boundary
conditions for the three-cluster continuum. This model  allowed us to
find a set of narrow and fairly wide resonance states in the three-cluster
continuum of $_{\Lambda}^{7}$Li. Analysis of resonance wave functions
reveals that the narrow resonance states are very compact three-cluster
configurations with small distances between interacting clusters. 

These results can be considered a prediction of the existence of narrow resonance
states in hypernucleus $_{\Lambda}^{7}$Li and can be used for planning of
future experiments.

\vskip3mm \textit{ 
	This work was partially supported by  the Science Committee of the
		Ministry of Education and Science of the Republic of Kazakhstan (Grant No.
		AP22683187, the project title "Structure of the light nuclei and hypernuclei
		in multi-channel and multi-cluster models") and by the Program of Fundamental
	Research of the Physics and Astronomy Department of the National Academy of
	Sciences of Ukraine (Project No. 0122U000889). V.V.S. is also grateful to the
	Simons Foundation for their financial support (Award ID: SFI-PD-Ukraine-00014580).}

\bibliographystyle{ieeetr}

\vspace*{-5mm} \rezume{%
	Н. Калжiгiтов, С. Амангельдінова, В.С. Василевський} {СТРУКТУРА ГІПЕРЯДРА $_{\Lambda}^{7}$Li В РАМКАХ МІКРОСКОПІЧНОЇ ТРИКЛАСТЕРНОЇ МОДЕЛІ} {Зв'язані та резонансні стани гіперядра $_{\Lambda}^{7}$Li досліджуються
	в рамках трикластерної моделі. Це ядро розглядається як трикластерна структура, що складається з $^{4}$He, дейтрона та лямбда-гіперона. Обрана трикластерна конфігурація дозволяє нам точніше описати структуру гіперядра $_{\Lambda}^{7}$Li та динаміку різних процесів, які включають взаємодію найлегших ядер та гіперядер. Головною метою даних досліджень є знаходження резонансних станів у трикластерному континуумі $_{\Lambda}^{7}$Li та визначення їхньої природи. У діапазоні енергій 0 $< E \leq$ 2 Мев вище трикластерного порогу $^{4}$He+$d$+$\Lambda$  виявлено низку вузьких резонансних станів.}{\textit{К\,л\,ю\,ч\,о\,в\,і\, с\,л\,о\,в\,а:} Кластерна модель, резонансні стани, трикластерна модель, гіперядра.}

\end{document}